\documentclass[11pt]{article}

\usepackage{graphics}   %---Comment out if we don't want PS support
\usepackage{endnotes}            %---Comment out if we don't want to be able to change footnotes to endnotes
\usepackage{amsfonts}            %---Comment if we don't want AMS fonts
\usepackage{amssymb}             %---Comment if we don't want AMS symbols
\usepackage{amsthm}              %---Comment if we don't want fancy theorem environments
\usepackage{amsmath}            %---Comment if we don't want various AMS math goodies
\usepackage{algorithm}           %---Comment if we don't want algorithm support
\usepackage{algorithmic}
\usepackage{soul}
\usepackage{blkarray}
\sethlcolor{black}
\usepackage{xcolor}
\makeatletter
\usepackage[letterpaper]{geometry}%---Uncomment if we want to allow paper size/margin changes
\usepackage{rotating}            %---Uncomment if we don't want to be able to rotate floats
\usepackage{color}
\usepackage{multibib}

\usepackage{setspace}                %---Comment to remove space-changing stuff
\usepackage{natbib}
\begin{document}

%---Document title page starts here
\title{Incorporating Structural Stigma into Network Analysis\thanks{This research was supported in part by ARO award W911NF-14-1-055 and NSF award SES-1826589.}
}
%To unblind just remove the comments
%--Author(s), date, etc.
\author{
Francis Lee\thanks{Department of Sociology, University of California, Irvine} and Carter T. Butts\thanks{Departments of Sociology, Statistics, Computer Science, EECS, and Institute for Mathematical Behavioral Sciences, University of California, Irvine}
}

\date{7/31/19}
\maketitle

%---Document abstract and keywords
\begin{abstract}
A rich literature has explored the modeling of homophily and other forms of nonuniform mixing associated with individual-level covariates within the exponential family random graph (ERGM) framework.  Such differential mixing does not fully explain phenomena such as stigma, however, which involve the active maintenance of social boundaries by ostracism of persons with out-group ties.  Here, we introduce a new statistic that allows for such effects to be captured, making it possible to probe for the potential presence of boundary maintenance above and beyond simple differences in nomination rates.  We demonstrate this statistic in the context of gender segregation in a school classroom.

\emph{Keywords:} social network analysis, homophily, stigma, social sanctions, xenophobia, ERGM
\end{abstract}

%---Definitions for Defs, Theorems, etc.
\theoremstyle{plain}                        %---Comment out this line if not using amsthm
\newtheorem{axiom}{Axiom}
\newtheorem{lemma}{Lemma}
\newtheorem{theorem}{Theorem}
\newtheorem{corollary}{Corollary}

\theoremstyle{definition}                 %---Comment out this line if not using amsthm
\newtheorem{definition}{Definition}
\newtheorem{hypothesis}{Hypothesis}
\newtheorem{conjecture}{Conjecture}
\newtheorem{example}{Example}

\theoremstyle{remark}                    %---Comment out this line if not using amsthm
\newtheorem{remark}{Remark}

%\ctb{You will note that, throughout, I've for the most part swapped out discrimination with segregation, stigma, etc.  That's because what is being modeled here is not really well-described as discrimination (which usually refers to the tendency to preferentially award benefits to members of one group rather than another, net of other factors).  Discrimination and segregation due to xenophobia can and do co-occur, but they aren't really the same thing.}

%\ctb{Refs should be added/updated.  I just stuck a few placeholders/mnemonics in for now.}

\section{Introduction}
%Exponential random graph models have helped to formalize understanding of social interaction. A long-standing issue in understanding social structure is how homophily arises. One particularly underexplored area is from tacit social knowledge, wherein friends will sanction or pressure friends from making friends to those who do not match up with some ascriptive attribute. While one could imagine this ascriptive attribute to either be latent, manifest.  
Homophily (the tendency for individuals to be tied to others with similar characteristics) and segregation (the tendency for individuals with different characteristics not to be in contact, nor to be found within the same social or geographical settings) are widely studied structural phenomena across a range of contexts \citep[see e.g.][]{blau1994structural, schelling1971dynamic, sakoda:jms:1971, schaefer2017friends, mcpherson2001birds, goodreau2017sources, butts:sm:2007b}. Within the specific context of cross-sectional social network models, these phenomena are generally treated within the rubric of \emph{differential mixing}, i.e. as a net tendency for ties to occur at higher or lower rates within pairs of individuals with particular combinations of attributes, controlling for other factors.  Mixing effects are easily incorporated into modeling frameworks such as the exponential family random graph models (ERGMs), and indeed such effects have been used to shed light on a variety of social phenomena including power in prison gangs \citep{schaefer2017friends}, and HIV diffusion \citep{goodreau2017sources}. 
%\ctb{Deleted various things that weren't necessarily ERGM related.  Be sure you are citing papers that actually use ERGMs, and not e.g. SAOMs!}.

While differential mixing is an important dimension of social structure, it does not by itself account for all aspects of segregation within social networks.  In particular, it does not capture the \emph{active maintenance of group boundaries} observed in many settings.  As \citet{goffman:bk:1963} famously noted, interacting with persons from other groups creates the risk that one's social identity will become ``spoiled'' via the association; that is, fellow in-group members may come to view ego as ``contaminated'' by associating with those having out-group membership, and their in-group status may become suspect.  This, in turn, may lead members of ego's in-group to withdraw contact from him or her.  As Goffman observes, this mechanism is often sufficiently well-understood that ego will anticipate becoming ostracized for out-group contacts, and simply avoid creating them in the first place (especially where ego's in-group contacts are numerous, and the costs of ostracism correspondingly high).
A similar phenomenon was posited by \citet{heider:jp:1946} in the context of balance theory.  Within an individual's mental model, group membership (a form of \emph{unit relation} in Heider's terminology) acts as an implicitly positive tie; if a perceiver, X, views another individual A as having a positive-valance relation to some individual B belonging to group G, then this therefore creates a positive two-path (A$\to$B, B$\to$G) in X's mental model.  Because closure of a positive two-path by a negative edge is unbalanced, X is predicted to perceive A as having a non-negative (and likely positive) association with G.  Now, consider the case in which X has a strongly negative association with G and sees the A$\to$G relation as positive.  For X to form a positive tie to A is now unbalanced, and hence dissonant.  Thus, the perception that someone is consorting with a member of a negatively perceived group can suppress willingness to interact with them positively, on purely affective (balance theoretic) grounds.  Where group members perceive their own groups positively and others negatively, this mechanism leads to ostracism of boundary spanners. 

While the social psychological mechanisms described by Goffman (stigma) and Heider (affective balance) are distinct, both have a common consequence for group structure: under the conditions where either is active, we should see that (ceteris paribus) (1) ego's propensity to form and maintain out-group ties is declining with their number of in-group ties; and (2) in-group members' propensity to form and main ties with ego is declining with ego's number of out-group ties.  In this way, both mechanisms serve to \emph{actively maintain group boundaries,} by discouraging boundary spanning and by marginalizing those who engage in it.  This is in contrast with standard differential mixing effects, which (in the case of homophilous mixing) simply posit a constant reduced baseline probability of cross-group versus within-group ties.  Homophilous mixing does not impose a conditional dependence between in-group and out-group ties, and does not per se lead to marginalization of boundary spanners.  As such, general tendencies towards homophilous mixing do not capture active boundary maintenance of the sort described here.

In the remainder of this paper, we introduce a simple family of ERGM statistics that better captures active boundary maintenance, either in the general case (in-group versus out-group) or in the special case of stigmatized groups (where interactions with members of a particular group are subject to sanction by non-members).  As we show, models with these statistics behave differently from models based on differential mixing alone, and their inclusion in an ERGM along with standard mixing terms can be used to test for the presence of boundary maintenance mechanisms.  We also illustrate the use of these terms with a case study involving Parker and Asher's (1993) study of classroom friendships, demonstrating that such mechanisms do appear to be active in some social networks. We conclude with the introduction of a framework for translating ERGM effects into ties gained or lost when the network is perturbed.

\section{Measuring Boundary Maintenance with Inhomogenous Star Statistics}

To develop a statistic for use in modeling boundary maintenance, we begin by assuming a set of $N$ actors, each of whom is a member of some specified group (denoted $A$, $B$, $C$, etc.).  We say that group $B$ is \emph{sigmatized} vis a vis the members of $A$ if the boundary between $A$ and $B$ is actively maintained by the members of $A$.  In the special case of homophilous interaction, each group is stigmatized vis a vis each other group (i.e., all boundaries are actively maintained).  From our above discussion, we begin with the basic intuition that, where $A$ actively maintains its boundary with $B$, the following should hold:
\begin{enumerate}
\item Let $i$ be a member of group $A$ and $j$ a member of $B$.   Then the conditional probability of an $i,j$ edge declines with the number of ties from members of $A$ to $i$.
\item Let $i$ and $j$ both be members of group $A$.  Then the conditional probability of an $i,j$ edge declines with the number of ties from $j$ to members of group $B$.
\end{enumerate}

To implement this intuition within an ERGM context, we recall that the conditional probability of an $i,j$ edge in a graph represented by adjacency matrix $Y$ is given by
\begin{equation}
\Pr(Y_{ij}=1|Y^c_{ij}=y^c_{ij},X,\theta) = \mathrm{logit}^{-1}\left(\theta^T \Delta_{i,j,t}(y,X)\right), \label{ergm_condprob}
\end{equation}
where $\mathrm{logit}^{-1}$ is the inverse logit function, $Y^c_{ij}$ is the set of all edge variables other than $Y_{ij}$, $\theta$ is a parameter vector, $t$ is a vector of sufficient statistics, and $X$ is a set of covariates.\footnote{For brevity, we have tacitly taken the reference measure to be constant, since this will not affect our development.}  $\Delta_{i,j,t}$ here is the \emph{changescore functional}, defined by
\begin{equation}
\Delta_{i,j,t}(y,X) = t(y^+_{ij},X)-t(y^-_{ij},X) \label{e_change}
\end{equation}
where $y^+_{ij}$ and $y^-_{ij}$ are matrices such that $y^+_{kl}=y^c_{kl}$ and $y^-_{kl}=y^c_{kl}$  for $(k,l)\neq(i,j)$ and otherwise $y^+_{ij}=1$, $y^-_{ij}=0$.  Intuitively, $\Delta_{i,j,t}(y,X)$ simply returns the difference in the vector of sufficient statistics, $t$, obtained by setting $y_{ij}$ to 1 versus 0, holding the rest of the network (i.e., $y^c_{ij}$) fixed.  Since the conditional edge probability is monotone in the changescore, it follows that our goal is to select a statistic such that the difference in the statistic associated with adding or removing an edge satisfies our substantive criteria.

To define such a statistic, we begin by letting $X$ be a vector of group memberships, such that $X_i$ indicates the group to which vertex $i\in V$ belongs (with $V$ being the vertex set).  As before, we let $A$ be a reference group, and $B$ another group that is stigmatized vis a vis the members of $A$.  We then define the following \emph{$AB$-inhomgeneous 2-star statistic}:
\begin{equation}
t_{AB}(y,X) = \sum_{i\in V} \sum_{j\in \{V\setminus i\}} \sum_{k\in \{V\setminus i,j\}} y_{ij} y_{jk} I(X_i=A) I(X_j=A) I(X_k = B), \label{e_stat_ab}
\end{equation}
where $I$ is an indicator for the truth value of its arguments. $t_{AB}$ counts the number of 2-stars whose first two vertices belong to $A$, and whose third vertex belongs to $B$ (i.e., that contain an edge from a member of $A$ to a boundary spanner with an edge to a member of $B$).  To see the effect of this statistic on edge probability, we compute the associated changescore function using Equation~\ref{e_change},
\begin{align}
\Delta_{i,j,t_{AB}}(y,X) &= t_{AB}(y^+_{ij},X)-t_{AB}(y^-_{ij},X)\\
&=\begin{cases} \sum_{k\in \{V\setminus i,j\}} y_{jk} I(X_k=B) & (X_i=A)\vee (X_j=A) \\ \sum_{k\in \{V\setminus i,j\}} y_{ki} I(X_k=A) & (X_i=A) \vee (X_j=B) \\ 0 & \mathrm{otherwise}\end{cases}.
\end{align}
Note that, when $i$ and $j$ are both in $A$, $\Delta_{i,j,t_{AB}}(y,X)$ counts the number of ties from $j$ to $B$.  Likewise, when $i$ belongs to $A$ and $j$ to members of $B$, $\Delta_{i,j,t_{AB}}(y,X)$ counts the number of ties from members of $A$ to $i$.  Thus, when the associated parameter $\theta_{AB}$ is negative, $t_{AB}$ reduces the conditional log odds of a tie from one member of $A$ to another by $|\theta_{AB}|$ for each tie that the second member of $A$ has to the stigmatized group.  Likewise, this same condition reduces the conditional log odds of a tie from a member of $A$ to the stigmatized group by $|\theta_{AB}|$ for each tie that the would-be sender receives from other members of $A$.  It follows, then, that $t_{AB}$ satisfies our desiderata for a statistic implementing active maintenance by group $A$ vis a vis a focal stigmatized group. 

\begin{figure}[h!]
	\begin{center}
		\includegraphics[height=2in]{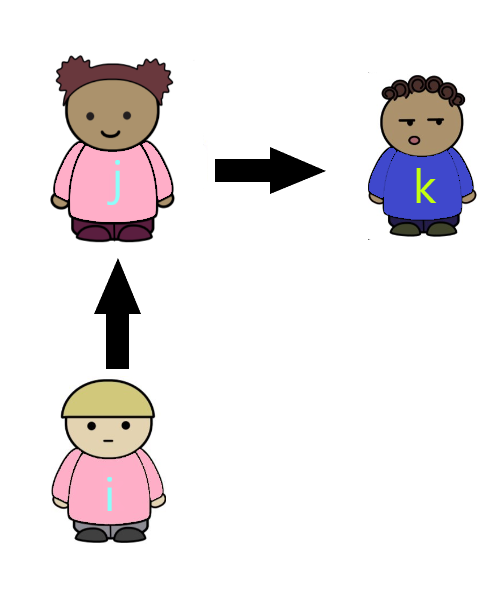}
	\caption{The above depicts an inhomogeneous 2-star, where the relevant group identity is determined based on the color of one's clothing. An actor $j$ wearing pink sends a cross-group tie to actor $k$ wearing blue. Actor $j$ also receives a tie from another actor $i$ wearing pink. \label{ihomo2star}}
	\end{center}
\end{figure}

Having treated the case of one group stigmatized vis a vis another, we can further generalize to other cases.  Arguably the most important case is that of homophilous interaction (here better thought of as \emph{xenophobia} in the structural sense used e.g. by \citet{butts:sm:2007b}), where each group is stigmatized vis a vis each other group.  To implement this effect, we propose a generalized \emph{inhomogeneous 2-star} statistic,
\begin{equation}
t_{I2P}(y,X) = \sum_{i\in V} \sum_{j\in \{V\setminus i\}} \sum_{k\in \{V\setminus i,j\}} y_{ij} y_{jk} I(X_i=X_j) I(X_j\neq X_k). \label{e_stat_i2p}
\end{equation}
As the name implies, $t_{I2S}$ counts the number of 2-stars with the central actor breaching their group boundary and forming a tie to another group. %that begin within one group and cross into another. 
To understand the action of $t_{I2S}$ on the graph, we consider its changescore function,
\begin{align}
\Delta_{i,j,t_{I2S}}(y,X) &= t_{I2S}(y^+_{ij},X)-t_{I2S}(y^-_{ij},X)\\
&=\begin{cases} \sum_{k\in \{V\setminus i,j\}} y_{jk} I(X_j\neq X_k) & X_i=X_j \\ \sum_{k\in \{V\setminus i,j\}} y_{ki} I(X_k=X_i) & X_i\neq X_j \\ 0 & \mathrm{otherwise}\end{cases},
\end{align}
which is clearly the number of outgoing ties from $j$ to out-group members (when $i$ and $j$ are in the same group), or the number of incoming ties from in-group members to $i$ (when $i$ and $j$ are in different groups).  As before, these have a suppressive effect on tie probability when the associated parameter, $\theta_{I2S}$, is negative, implementing the notion that in-group ties reduce the propensity to bridge to out-group members, and (likewise) that out-group ties reduce the propensity to be selected by in-group members.  Inclusion of $t_{I2S}$ hence provides a way of modeling active boundary maintenance in the context of homophilous interaction. A minimal example of an inhomogeneous 2-star is shown graphically in Figure~\ref{ihomo2star}.

\section{Simulation Study}

To demonstrate the capacity of the inhomogeneous two-star statistic to implement group boundary maintenance, we simulate group structure under three scenarios.  One is a simple model without active boundary maintenance, including only edge and node mixing terms.  The second adds a boundary maintenance by incorporating an additional inhomogeneous 2-star effect. The third increases the tendencies to mix within group and penalizes a tendency to mix between groups as well as the inhomogeneous 2-star effect. %The fourth intensifies the tendencies to mix within group and penalizes a tendency to mix between groups even more, as well as incorporates the penalized inhomogeneous 2-star effect.  
We simulated networks using the \textit{ergm} package in R 3.5.0 \citep{hunter.et.al:jss:2008}, with a custom user extension from \textit{ergm.userterms} \citep{hunter.et.al:jss:2013}. Networks consisted of 2 group identities, each assigned to 20 individuals. Networks were simulated using the tie-no tie proposal, and with a thinning interval of 1000. 1000 networks were simulated for each parameter set. The simulation coefficients are presented below in Table~\ref{parameter1}.

\begin{table}[H]
	\centering
	\begin{tabular}{rrrrrr}
		\hline
		& Edges & Nodematch & Nodemix (Between Groups) & Inhomogeneous 2-Star \\ 
		\hline
		Model 1 &-2 & .75 & -2.5  & 0 \\ 
		Model 2 & -2 & .75 & -2.5 & (-1,-0.95,...,0) \\ 
		Model 3 & -2 & 3 & -1 & (-1,-0.95,...,0) \\ 
		\hline
	\end{tabular}
	\label{parameter1}
	\caption{Parameters for the simulation experiment. Models 2 and 3 incorporate an inhomogeneous 2-star term, whose coefficient varies from -1 to 0 in 0.05 increments.}
\end{table}

%\begin{table}[H]
%	\centering
%	\begin{tabular}{rrrrrr}
%		\hline
%		& Edges & Nodematch & Nodemix (Between Groups) & Inhomogeneous 2-Star \\ 
%		\hline
%		Model 1 &-2 & .75 & -2.5  & 0 \\ 
%		Model 2 & -2 & .75 & -2.5 & (-1,-0.95,...,0) \\ 
%		Model 4 & -2 & 2 & -.75 & (-1,-0.95,...,0) \\
%		Model 3 & -2 & 3 & -1 & (-1,-0.95,...,0) \\ 
%		\hline
%	\end{tabular}
%\label{parameter}
%\caption{This table depicts the simulation coefficients. Models 2, 3, and 4 incorporate the inhomogeneous 2-star term whose coefficient is represented by i, which will vary from -1 to 0 in 0.05 increments. The lightly colored regions depict the 95\% simulations intervals.}
%\end{table}

\begin{figure}[H]
	\begin{center}
		\includegraphics[width=4in]{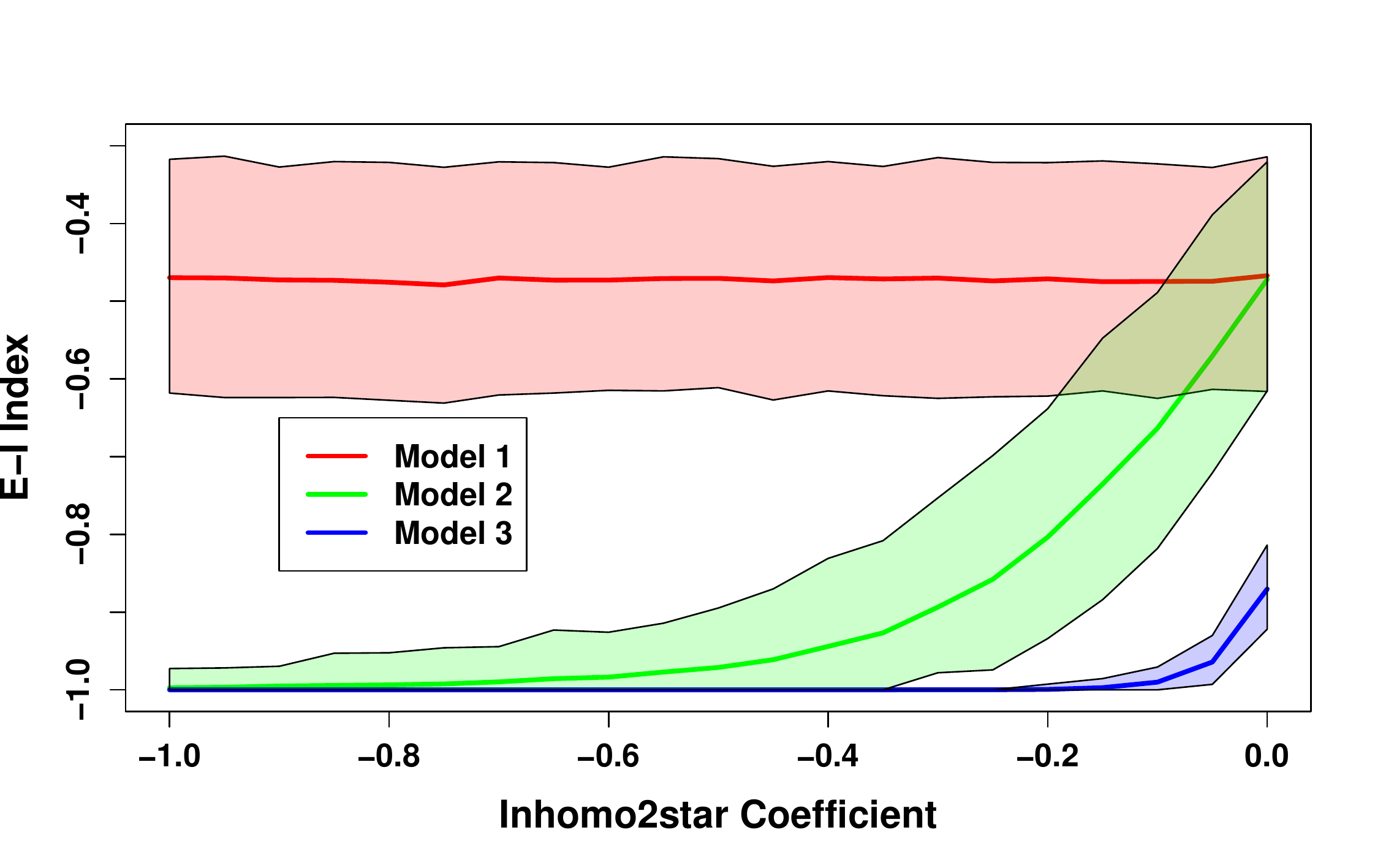}
	\caption{The mean E-I Index of the three simulations (solid lines) and 95\% simulation intervals. Model 1 (red) includes edges and node mixing terms only. Model 2 (green) incorporates the inhomogeneous 2-star alongside the parameters of the first model. Model 3 (blue) intensifies this tendency of mixing towards one's own group and disassociation with the other group. \label{label2}}
	\end{center}
\end{figure}

From our table of simulation parameters, we can see that in Model 1, the log odds of a tie existing between two individuals who are part of different groups will be -2.5. In Model 2, the log odds of a tie existing between two individuals who are part of different groups will be -2.5 plus the inhomogeneous 2-star coefficient times the number of within-group ties Individual 1 has plus the inhomogeneous 2-star coefficient times the number of within-group ties Individual 2 has. As the inhomogeneous 2-star coefficient varies from -1 to 0, this will penalize the log-odds by the number of within-group ties that they have. We examine the degree of group separation in each case by means of the E-I index, which measures the tendency for ties to be made between versus within groups  \citep{krackhardt.stern:spq:1988}. An E-I index of 1 indicates all ties are between groups, -1 indicates that all ties within group, and 0 indicates exact equality of between versus within group ties. We simulate 1000 networks in each condition, and calculate the mean E-I index.  The resulting E-I index values are shown in Figure~\ref{label2}. 

%\begin{figure}[h!]
%	\begin{center}
%		\includegraphics[width=4in]{./EIIndex.pdf}
%	\end{center}
%	\label{label1}
%	\caption{The mean E-I Index of the four simulations. Red depicts the edges and node mixing terms only. Green incorporates the inhomogeneous 2-star alongside the parameters of the first model. Blue increases this tendency of mixing towards one's own group and disassocation with the other group. Black increases the tendency even more.}
%\end{figure}

First, we note that the inclusion of active boundary maintenance effects have a substantial impact on group segregation, above and beyond what is produced by mixing effects.  We also observe that as we vary the inhomogeneous 2-star's negative effect, we see a relatively smooth segregation response (with segregation increasing rapidly as inhomogeneous two-stars are suppressed).  Moreover, in the strongly xenophobic model, we observe that the convergence to full segregation occurs very quickly as the parameter falls. Furthermore, the simulation intervals show some variation in the E-I index even with a large penalty to inhomogeneous 2-stars in the moderately segregated mixing regime, whereas in the strongly segregated regime, the variation surrounding the confidence interval is sharply suppressed until the inhomogeneous 2-star penalty decays in strength. This shows that active boundary maintenance \emph{amplifies and reinforces} mixing effects, making its impact on segregation larger when there is already some degree of mixing inhibition on which it can act.

\section{Empirical Example}

To show the empirical utility of the inhomogeneous two-star term for measuring stigmatization, we examine the \citet{parker1993friendship} friendship dataset, wherein a group of fifth grade students were asked to list the friends they had within their class. Within the network, there were 22 children, with 13 boys and 9 girls. Coding by gender, we see that girls and boys both have a tendency to homophilously associate, but the presence of the one boy who is only friends with girls and receives no ties from boys suggests that a stigmatic process may be at play: such a pattern would be exceedingly unlikely to occur from simple mixing effects. To more formally investigate this possibility, we proceed by fitting an ERGM with inhomogeneous 2-star effects. 

\begin{figure}[h!]
	\begin{center}
		\includegraphics[trim={0 3in 0 3in},width=3in,clip]{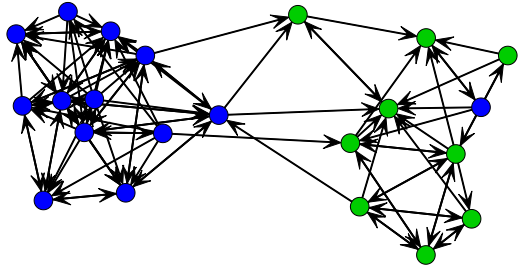}
	\caption{Friendship nominations among fifth-graders from Parker and Asher; blue nodes are boys, and green nodes are girls. The strong pattern of segregation, together with the observation that pupils with more cross-group ties tend to have fewer within-group ties, suggests active boundary maintenance by gender. \label{Friendnet}}
	\end{center}
\end{figure}

We model friendship nominations as follows. Friendship networks are well known to be driven by strong norms of reciprocity (mutual tie nomination) and social phenomena such as transitive friendship nomination also feature heavily in these networks. We can model these phenomena with statistics that calculate the number of reciprocal ties and the edgewise shared partner distribution. Qualitatively, we observe from Figure~\ref{Friendnet} that friendship nominations seem to be very obviously segregated within gender, except for a lone boy who has no male friends and multiple female friends. If the second-order process of stigma by association were absent, it would be extremely unlikely that we would simultaneously see so few cross-group ties not involving this student and so many cross-group ties involving this student, nor the coincidence of this anomaly with the student's being ostracized from his own group. By contrast, this is precisely the type of anomaly that would be expected under active boundary maintenance. 

\begin{figure}[h!]
\begin{center}
\includegraphics[width=2in]{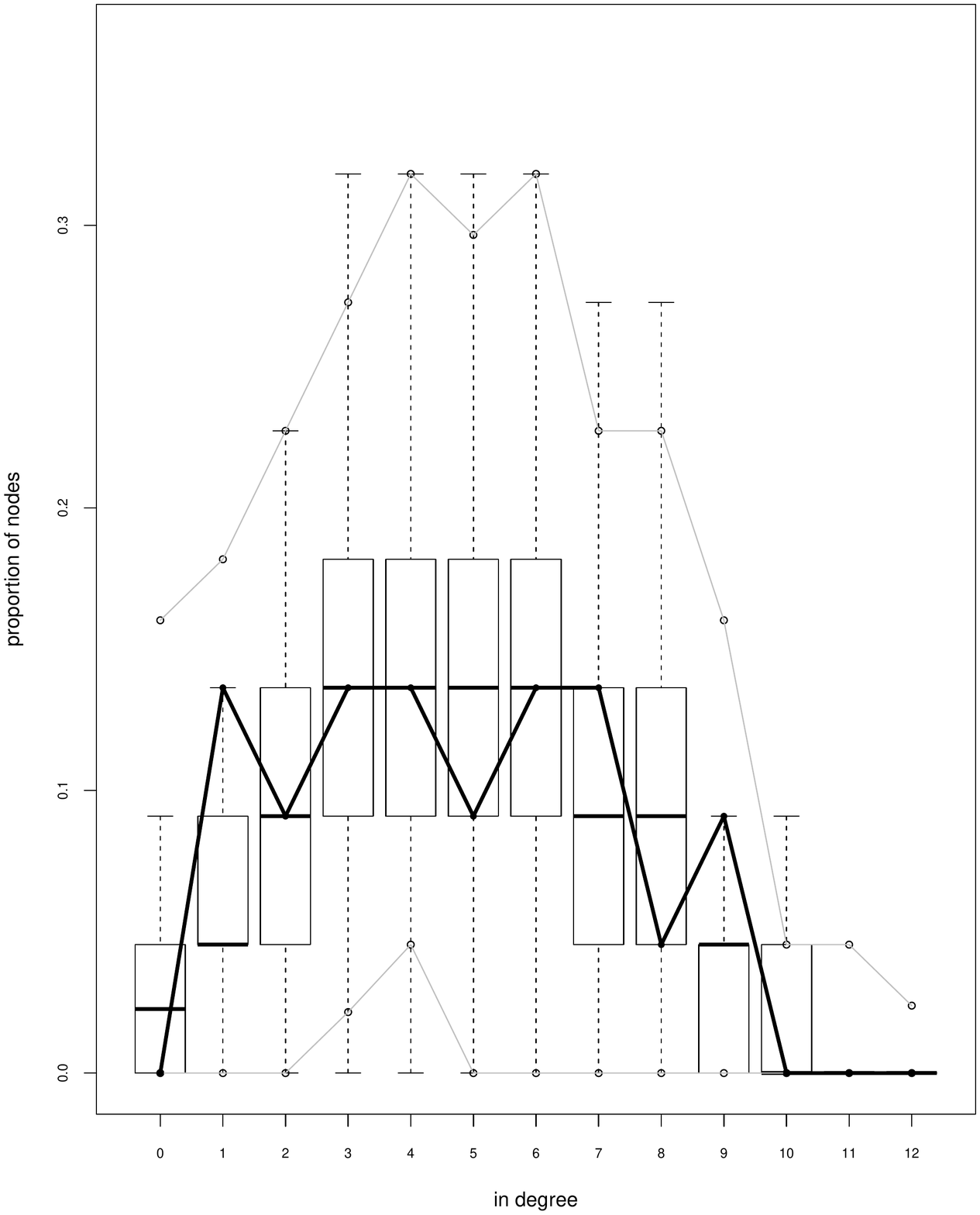}\includegraphics[width=2in]{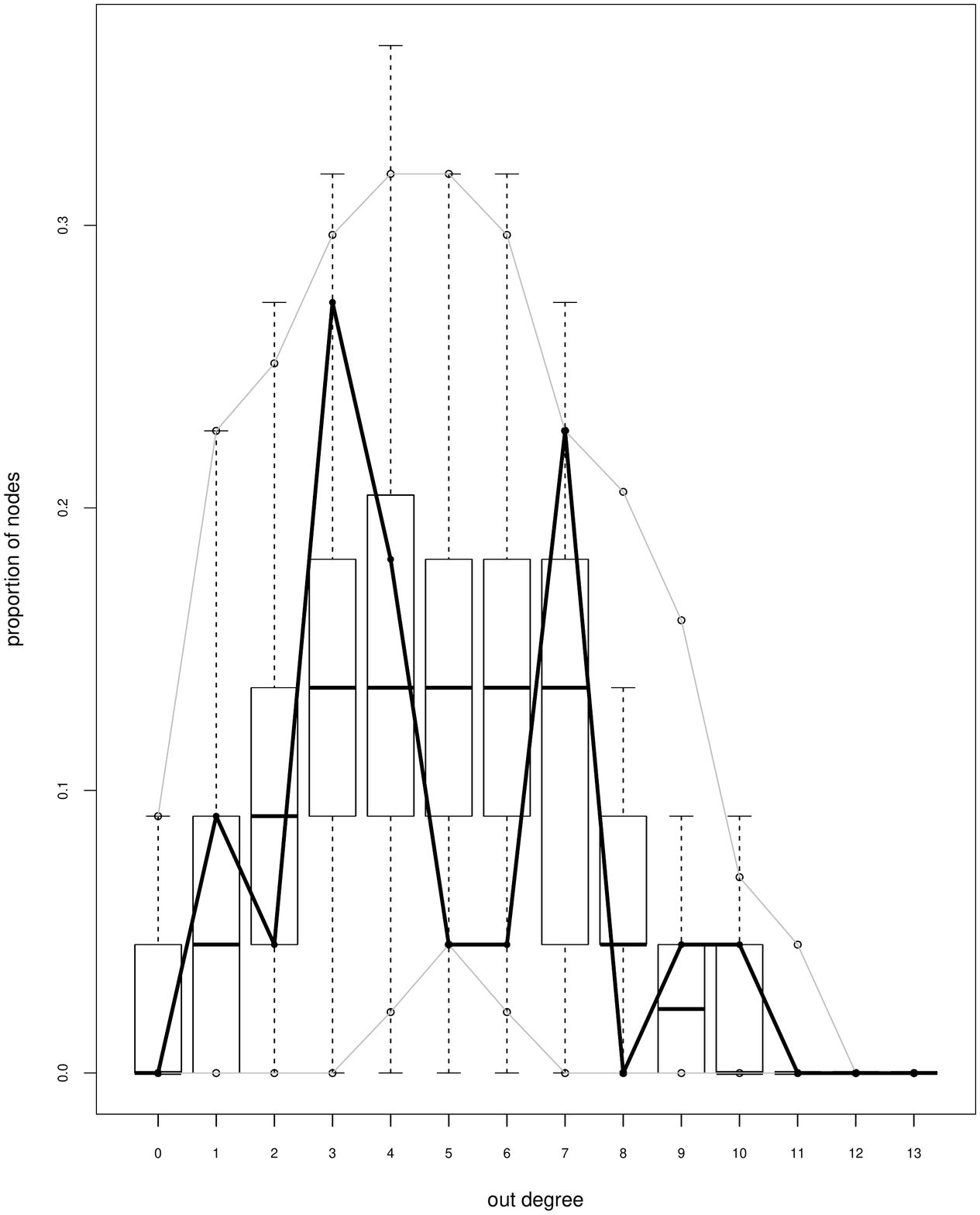}\\
\includegraphics[width=2in]{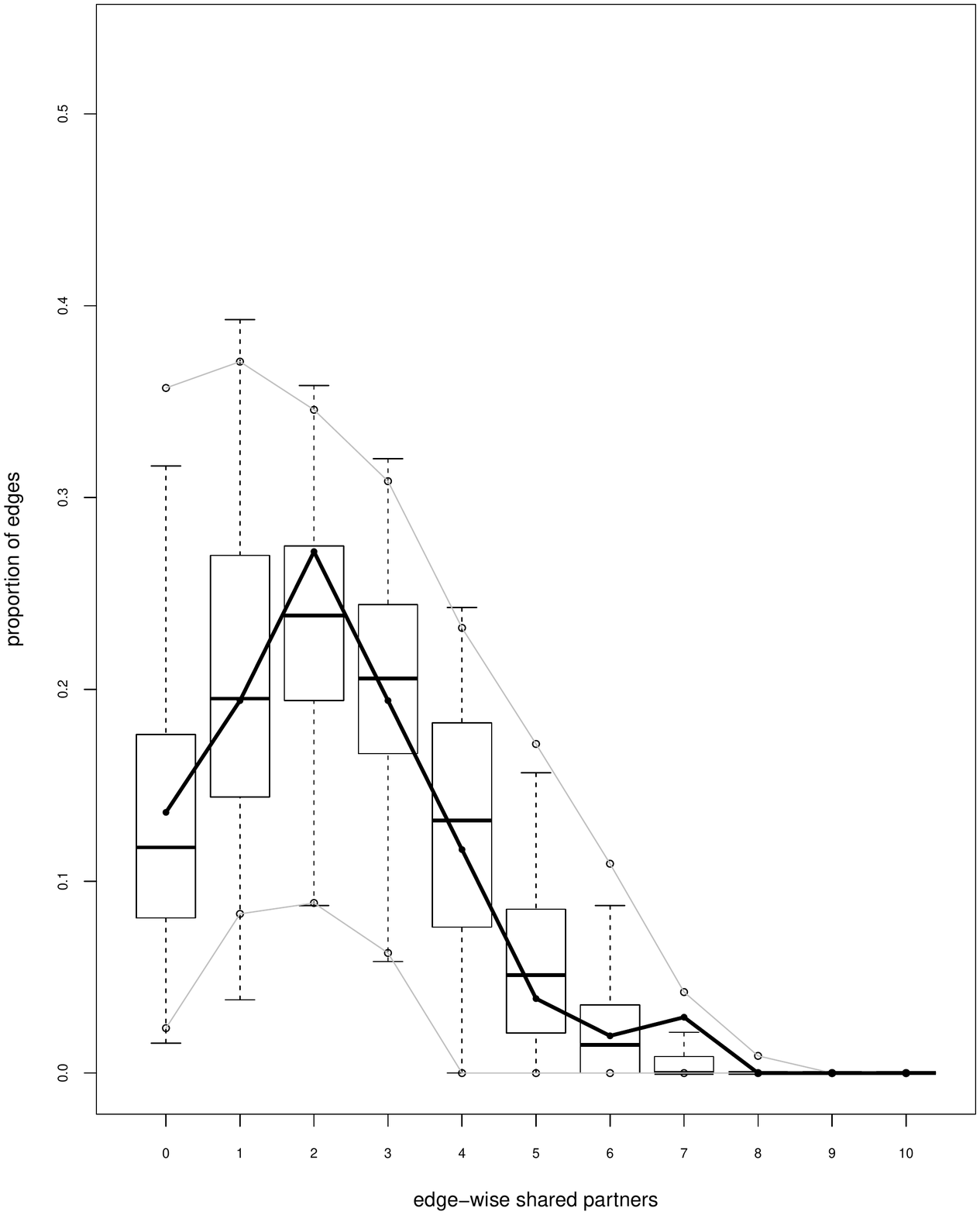}\includegraphics[width=2in]{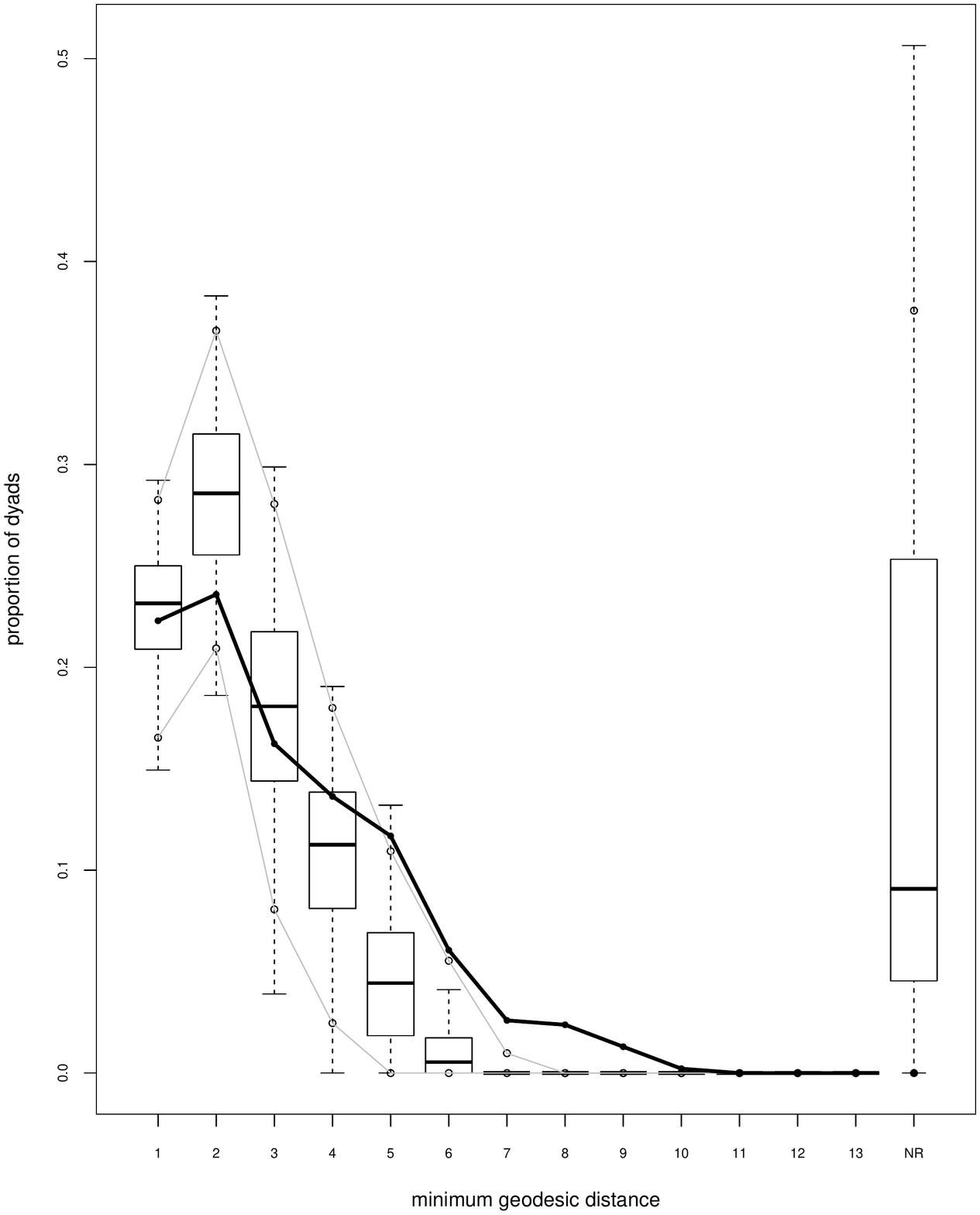}\\
\includegraphics[width=2in]{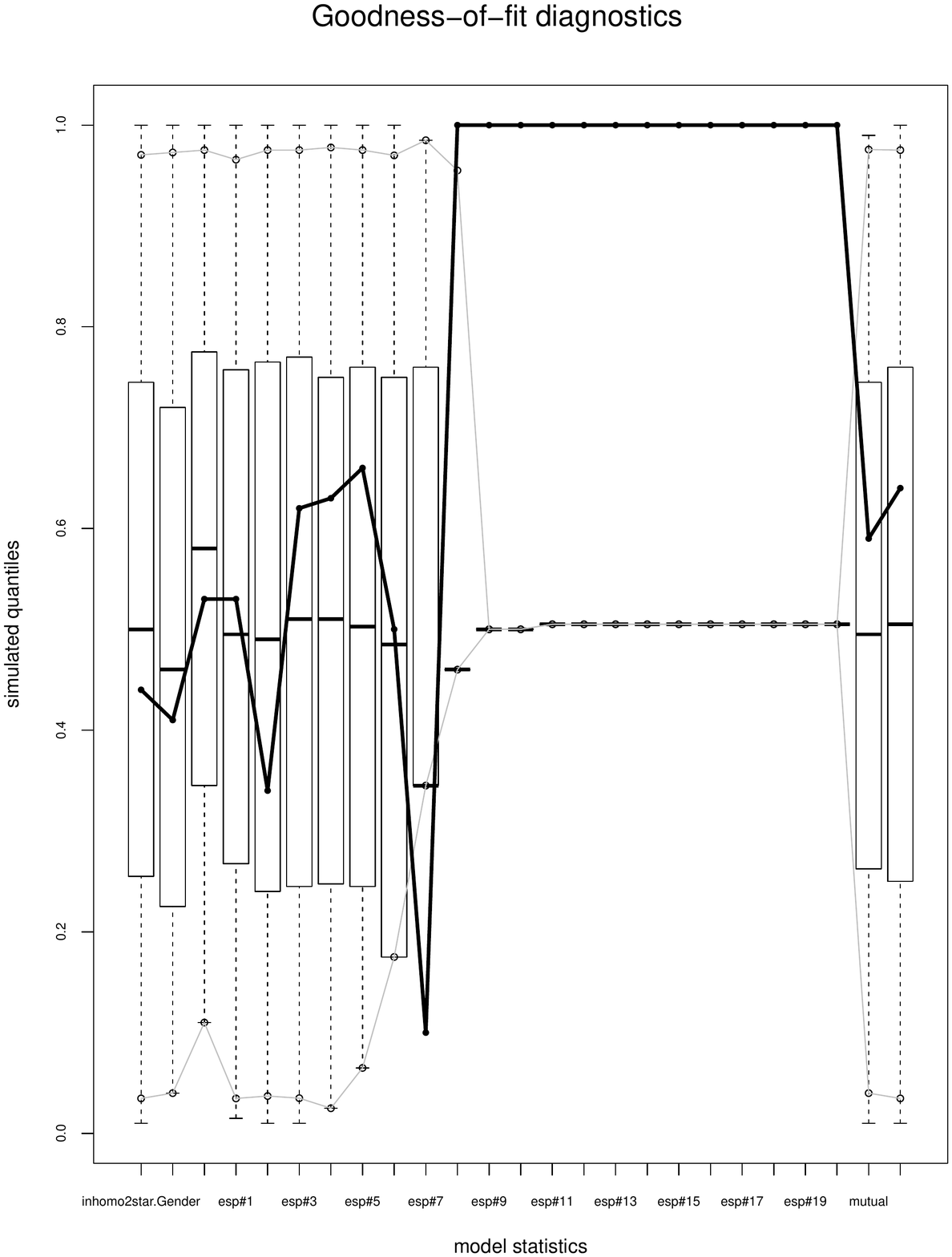}
\caption{The plot above depicts the various model adequacy checks. By simulating networks from the estimated model, we test whether various aspects of the simulated networks (e.g. the in/out-degree distribution, the geodesic distribution, the shared partner distribution, and the mean statistics) could potentially reproduce those same aspects observed in the dataset. The black lines depict the observed network statistics from the dataset, and the gray lines depict 95\% simulation intervals under the estimated model. \label{fit}}
\end{center}
\end{figure}

To test for this possibility, we include an inhomogeneous two-star term by gender with a baseline model including within and between group mixing, reciprocity, and an effect for triadic closure (GWESP).  As AIC comparison of this model versus a model with distinct within-group mixing rates by gender favored homogeneity, we employ a single base rate for within-group ties.  (Note that we do not require an explicit edge term, as the baseline tie probability across groups is set by nodemix terms, and the (homogeneous) within-group probability is set by a nodematch term.) In our model exploration process, we used best subset selection to exhaustively search what subset of model terms would generate models with the lowest AIC. The model terms we assessed were nodemix, globalized inhomogeneous 2-star, edges, mutuality, geometrically weighted in-degree, geometrically weighted out-degree, geometrically weighted edgewise shared partnership distribution, and geometrically weighted dyadwise shared partnership distribution. We then assessed the model adequacy of the 50 models with the lowest AIC. The selected model has the lowest AIC and passes ERGM model adequacy checks as shown in Figure~\ref{fit}. We present the coefficients below in Table ~\ref{Table1}. Models were estimated using the \textit{statnet} package in R 3.5.0 \citep{handcock.et.al:jss:2008}. The MCMC sample size was 10,000 and the MCMC interval was also 10,000.

\begin{table}[ht]
	\centering
	\begin{tabular}{rrrr}
		\hline
		& Estimate & Std. Error &  Pr($>|z|$) \\ 
		\hline
		Inhomo2star.Gender & -0.386 & 0.117 & 0.00095 \\ 
		Nodemix.Boys.Girls & -1.648 & 0.435 & $<1e-5$  \\ 
		Nodemix.Girls.Boys & -3.495 & 0.726 & $<1e-5$  \\ 
		GWESP & 0.475 & 0.225 & 0.035 \\ 
		GWESP ($\alpha$-decay) & 0.803 & 0.331 & 0.0153 \\ 
		Mutuality & 1.102 & 0.418 & 0.00846 \\ 
		Nodematch.Gender & -1.948 & 0.389 & $<1e-5$  \\ 
		\hline
	\end{tabular}
\label{Table1}
\caption{Estimated coefficients for friendship nominations. The null deviance is 640.5 with 462 degrees of freedom and the residual deviance of this model is 346.5 with 455 degrees of freedom.}
\end{table}

%\begin{table}[ht]
%\centering
%\begin{tabular}{rrrrr}
%  \hline
% & Estimate & Std. Error & MCMC \% & p-value \\ 
%  \hline
%edges & -2.6569 & 0.2706 & 0.0000 & $<$1e-04 \\ 
%  inhomo2star.Sex & -0.0725 & 0.0150 & 0.0000 & $<$1e-04 \\ 
%  gwesp.OTP & 0.2580 & 0.0245 & 0.0000 & $<$1e-04 \\ 
%  gwesp.OTP.decay & 2.0358 & 0.0324 & 0.0000 & $<$1e-04 \\ 
%  mutual & 1.2254 & 0.2571 & 0.0000 & $<$1e-04 \\ 
%  nodematch.Sex & 0.2201 & 0.1317 & 0.0000 & 0.0951 \\ 
%   \hline
%\end{tabular}
%\label{Table1}
%\caption{This describes the coefficients fit in our final ERG model.}
%\end{table}

The fitted model may be interpreted as follows.  The base log-odds of same-gender nominations is approximately -1.9, with boy-girl nominations having a nominally (but not significantly) higher log-odds of -1.6 and girl-boy nominations having a substantially lower base log-odds of -3.5.  We thus see an overall gender asymmetry in nominations, with boys tending to treat girls similarly to other boys, and girls tending to treat boys as substantially less attractive nomination targets than other girls (ceteris paribus).  A strong and significant reciprocity effect is present, although the estimated effect is not large enough to result in even odds of reciprocation holding out other factors.  A positive GWESP term indicates a tendency towards transitive closure, with the fairly large decay parameter (apx 0.8) indicating that the impact of multiple shared partners on closure probability is substantial.  All of the above are fairly typical of homophilous friendship networks.  With respect to the inhomogeneous two-star, we see a significant negative effect (apx -0.4, $p<0.001$), indicating the presence of active boundary maintenance.  Intuitively, this implies that (ceteris paribus) \emph{each} cross-gender tie sent by a student, $j$, decreases the conditional log-odds of an $(i,j)$ nomination from a student $i$ of the same gender by roughly 0.4; likewise, each incoming $(i,j)$ nomination decreases the conditional log-odds of a cross-gender tie sent by $j$ by the same margin.  Thus, even a small degree of group embeddedness is predicted to inhibit cross-group nominations, and even fairly minimal extension of cross-group ties is predicted to lead to substantial ostracism by one's own group.  We explore this phenomenon further in the next section, by means of conditional simulation.

\section{Investigating Boundary Maintenance via Network Perturbation}

To provide another way of looking at how boundary maintenance operates, we use the generative properties of the ERGM family to perform a \emph{perturbation analysis} of the Parker and Asher model, in effect asking how particular changes to the network - such as the creation of a cross-group tie - would be predicted to affect the conditional probability of observing other network features.  Specifically, we focus on two types of hypothetical scenarios which we find fruitful for understanding what the social forces found above mean for friendship networks. Each represents, in a basic sense, a violation of group boundary norms; the question is then how such a perturbation would be predicted to propagate to others' relationships with those involved in the violation, all other things being held constant.  The scenarios we consider are as follows:
\begin{enumerate}
	\item The tie variable $y_{AB}$ is toggled ``on'' (In this case, $A$ initiates a cross-group tie to non-neighbor $B$) 
	\item An exogenous covariate $x_A$ for actor $A$ changes value.  
\end{enumerate}

We note that one value of computational modeling is that we can examine unusual scenarios in terms of their impact on an identified set of social mechanisms; thus, while gender change (scenario two) would be unusual in this population, and would be accompanied in an actual event by other social consequences, here we can use a hypothetical gender change that involves \emph{only} the specified mechanisms as a tool to probe the effect of active boundary maintenance on friendship.  Both scenarios should be viewed as thought experiments rather than actionable predictions, though they shed some light on what would be expected to happen were no other mechanisms operational.

\subsubsection*{Scenario 1: $y_{AB}$ is toggled on}
\begin{figure}[H]
	\begin{center}
		\includegraphics[width=2in]{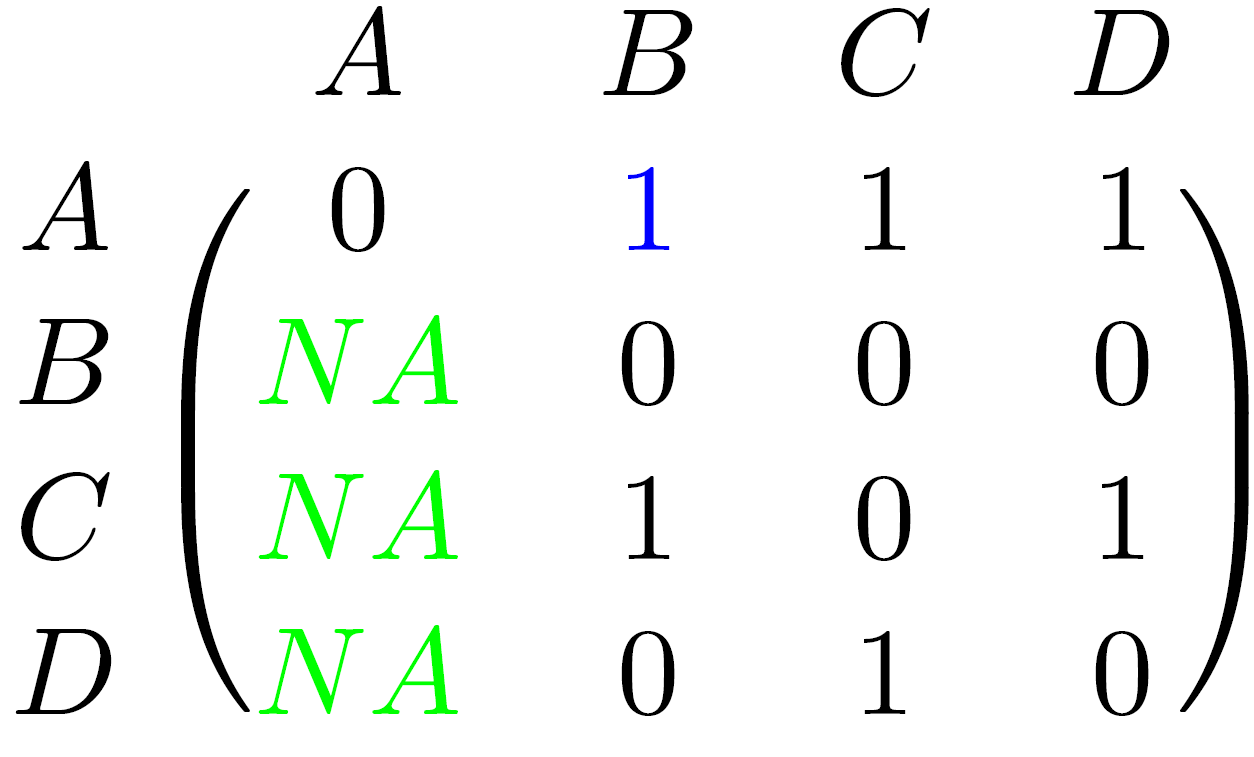}
\caption{Sociomatrix representation of the assessment of the expected change in in-degree as a result of ego $A$'s sending a cross-group tie to $B$ to whom they were not previously tied. \label{DeviantTieBlockArray}}
	\end{center}
\end{figure}

Although one could imagine wanting to examine the change in any network statistic associated with an exogenous tie change, an intuitive quantity that one may wish to examine (especially given the setting of stigma) is the expected change in the in-degree of $A$. That is, given that we add a tie from $A$ to out-group member $B$, what are the social consequences for person $A$ (ceteris paribus) in terms of incoming friendship nominations?  Specifically, we examine the difference
\begin{equation}
\mathbf{E}[\sum_j^J Y_{jA}|\theta,X,Y_{.A}^C=(y_{AB}^{+})^C_{.A}]-\mathbf{E}[\sum_j^J Y_{jA}|\theta,X,Y_{.A}^C=(y_{AB}^{-})^C_{. A}],
\end{equation}
\noindent i.e. the difference in the conditional expectation of $A$'s indegree when the new tie is present, versus when it is absent (holding the rest of the network constant), under our friendship model.  The coefficients $\theta$ are derived from our above-fit friendship model, and the exogenous covariates $X$ and the observed network $y$ come from the dataset. $y_{AB}$ refers to the tie variable being sent by actor $A$ to actor $B$. $(y_{AB}^+)^C$ and $(y_{AB}^-)^C$ refer to every single tie variable being fixed except for $y_{AB}$  which is either toggled on or off, respectively, and $y^C_{.A}$ refers to all elements of $y$ except the incoming ties to $A$. We represent this graphically in Figure~\ref{DeviantTieBlockArray}, we may observe the toggled tie is blue, and the green boxes to represent the focal examining the conditional probability of that tie. Everything colored black is fixed. The green values are the quantity of interest and are treated as random (here represented as ``missing''). We evaluate this quantity as the expected in-degree change, averaging over all possible realizations of the vector of incoming ties.

\begin{figure}[H]
 	\begin{center}
 		\includegraphics[width=5in]{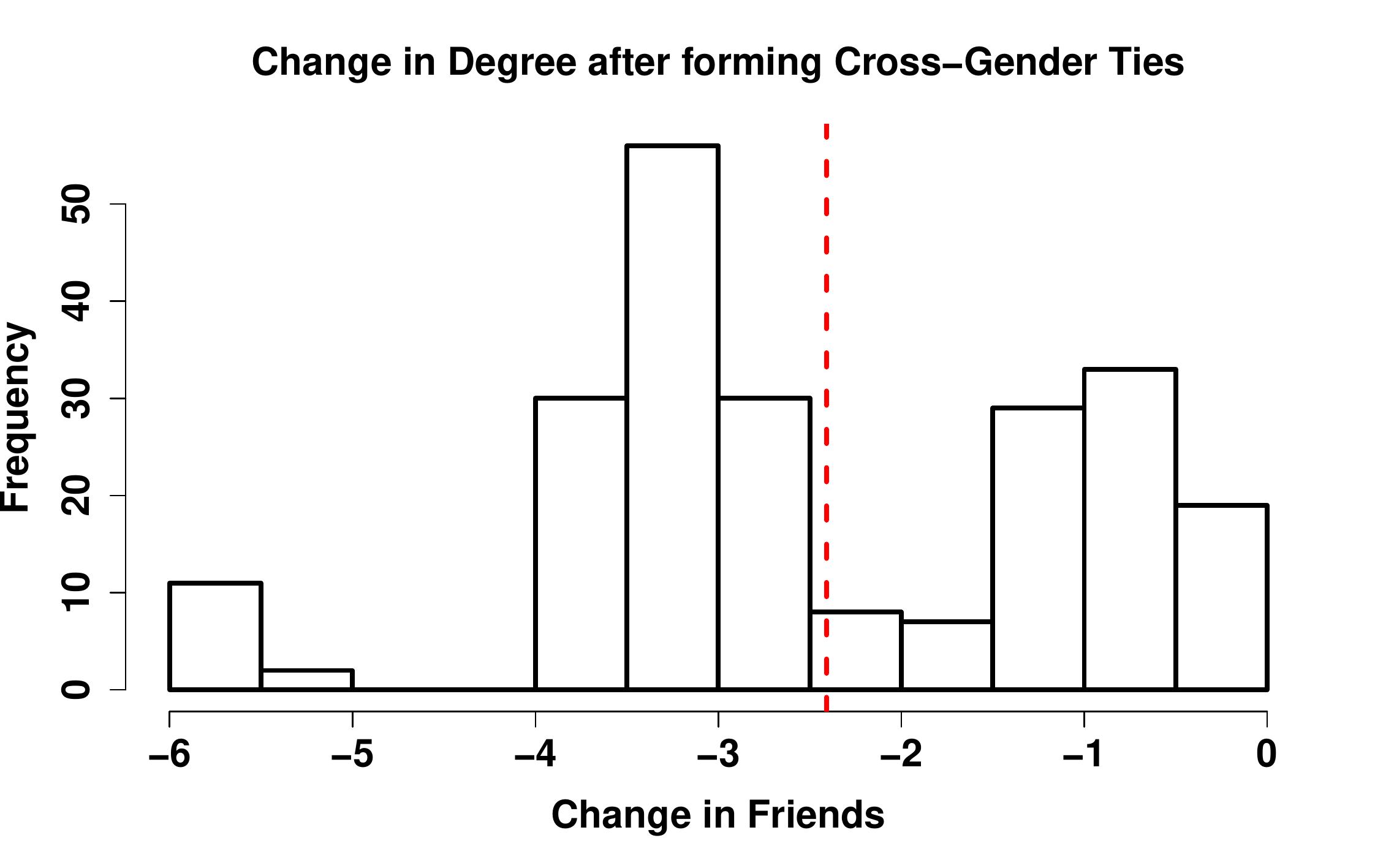}
 		\caption{Distribution of in-degree change for a random actor $A$, conditional on the rest of the network being at its observed state and a cross-gender tie added from $A$ to random individual $B$.  The dotted red line depicts the mean change in degree.  On average, forming a cross-group tie is predicted to result in loss of incoming friendship nominations. \label{DeviantTie}}
 	\end{center}
 \end{figure}

Within the context of the Parker and Asher dataset, we show the distribution of expected degree change when a cross-group tie is made in Figure~\ref{DeviantTie}.

As we can see in this scenario, sending a new cross-gender tie to a random member of the other group results in an average loss of 2.41 friendship nominations. Given that the average number of friendship nominations is 4.682, in expectation, an individual who forms a cross-gender tie can expect to lose roughly 51\% of their friendship nominations.  This outcome results despite the effects of reciprocity and GWESP, which tend to add edges in response to additional edges; here, these are outweighed by the boundary maintenance mechanism (implemented via the inhomogeneous two-star effect), which makes $A$ less attractive to their in-group when they send more ties to outsiders.

\subsubsection*{Scenario 2: $x_{A}$ for an actor $A$ changes value}
\begin{figure}[H]
	\begin{center}
		\includegraphics[width=2in]{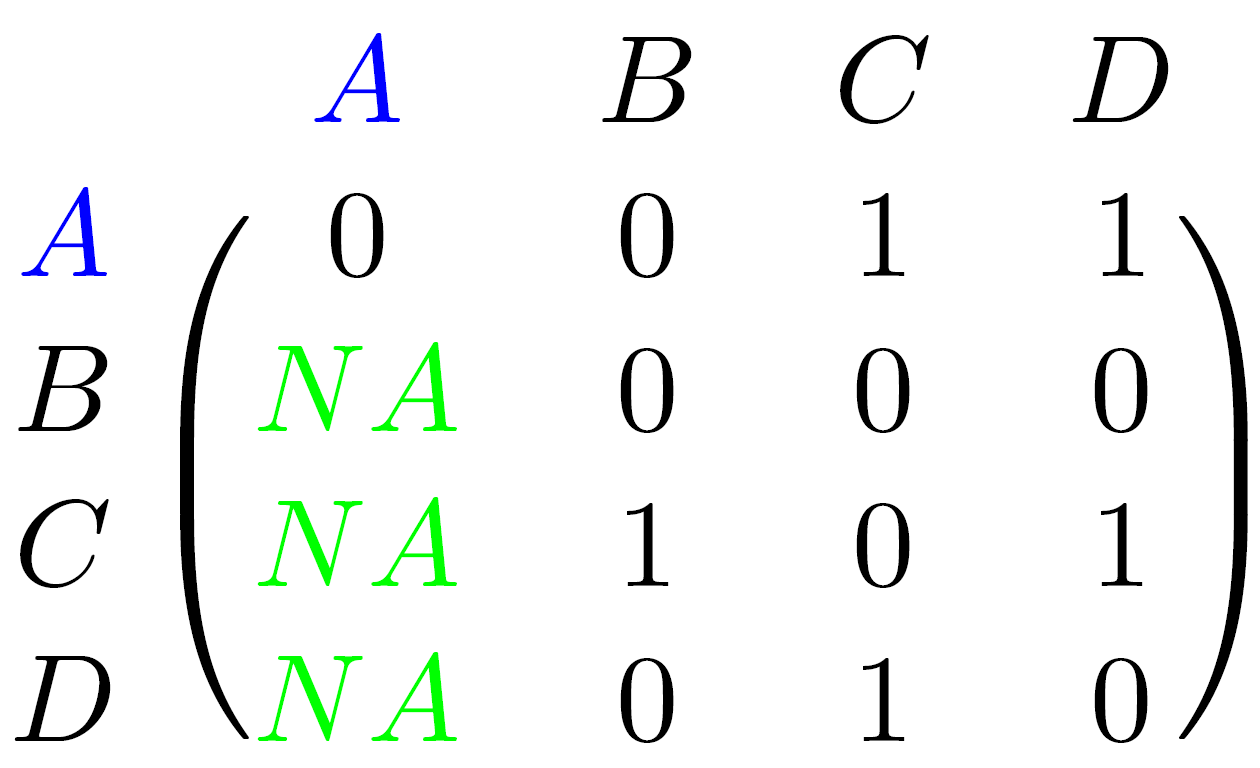}
\caption{Sociomatrix representation of the assessment of the expected change in in-degree as a result of ego $A$'s covariate changing. \label{CovChange}}
	\end{center}
\end{figure}

Similar to the above, we may observe the predicted consequence for $A$'s incoming friendship nominations when his or her membership status is changed, holding all else fixed. Though this can be viewed as an abstract thought experiment, it also provides a simplified scenario for what might happen when an individual previously thought to have a ''normal'' identity \citep{goffman:bk:1963} is revealed to have a stigmatized identity (e.g. HIV stigma, sexual orientation, gender orientation, criminal background).  In the context of the student network, gender change involves $A$'s switching to a stigmatized group from the perspective of $A$'s former in-group members, but at the same time switching away from a stigmatized group from the perspective of their former out-group members.  
The equation below describes the expected in-degree change of $A$ conditional on $A$'s changing exogenous covariate and all of the other ties remaining fixed:
\begin{equation}
E[\sum_j^J Y_{jA}|\theta,X',Y_{.A}^C=y^C_{.A}]-E[\sum_j^J Y_{jA}|\theta,X,Y_{.A}^C=y^C_{.A}],
\end{equation}
\noindent with $X_i'=X_i$ for all $i\neq A$, and $X'_A$ being ``flipped'' to the alternate state.  We represent this graphically in Figure~\ref{CovChange}.  Every color has the same meaning as in scenario 1, but in this case the external covariate on $A$ is set by the researcher to its new value.

 \begin{figure}[H]
	\begin{center}
		\includegraphics[width=5in]{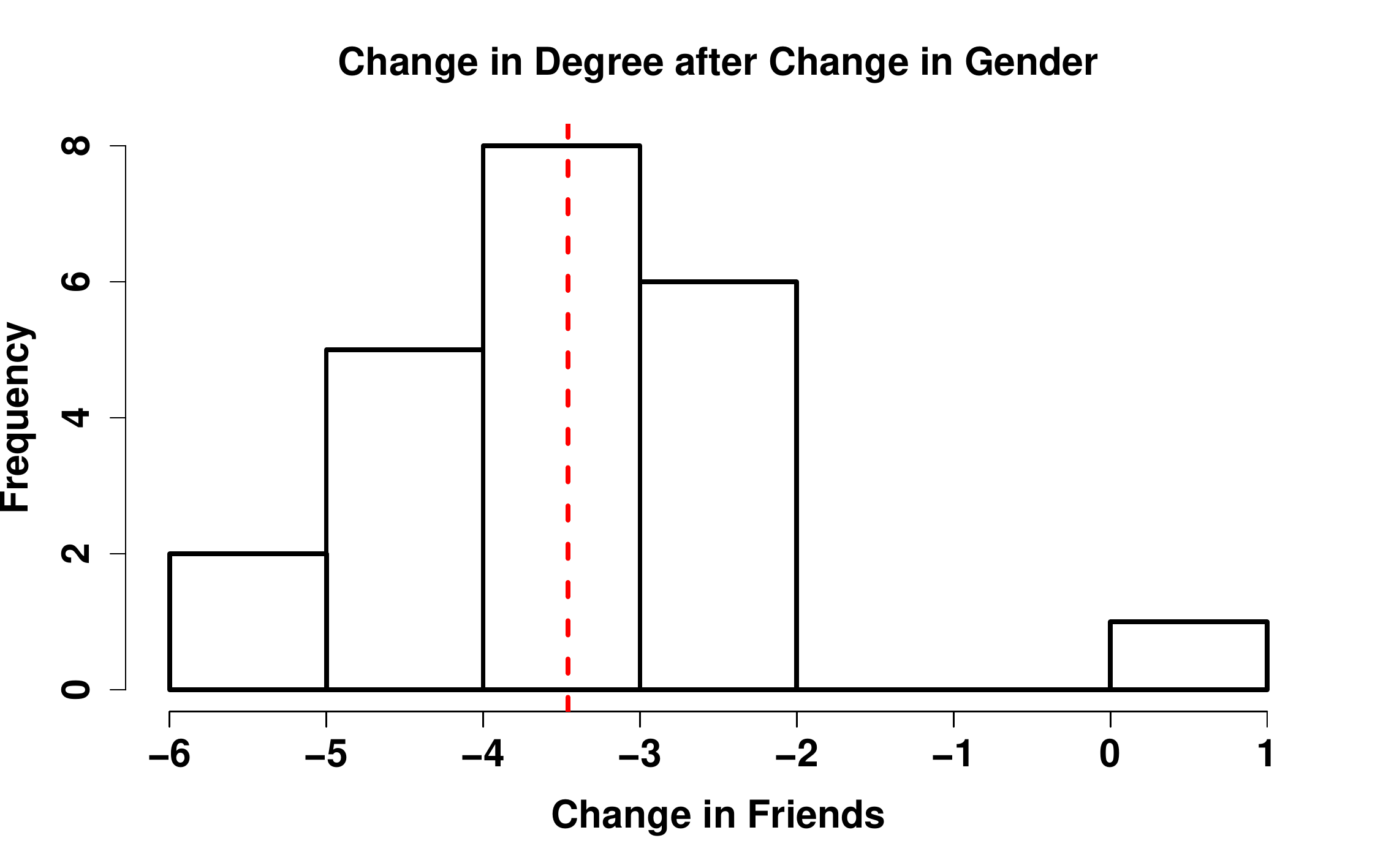}
		\caption{Distribution of in-degree change conditional on the rest of the network being constant when a random individual's gender is changed. The dotted red line depicts the mean change in in-degree.  On average, swapping group membership leads to tie loss. \label{ChangeinGender}}
	\end{center}
\end{figure}
We can see from Figure~\ref{ChangeinGender} that, on average, switching a random student's gender leads to a loss of 3.45 friendship nominations (approximately 73\%). The core driving mechanism here is the sudden reconsideration of ego's formerly within-gender nominations as cross-gender ties, making ego an unattractive nomination target for members of his or her new in-group; although reciprocity and closure effects involving ties to and among ego's former in-group will, on average, help preserve some nominations from that group, this is not sufficient to overcome the lack of nominations from the new in-group.  While actual gender transitions are far more complex than treated here, the simulation highlights a general challenge for actors crossing actively maintained boundaries: the same ties that once bound them to their in-group become liabilities to acceptance once a transition is made, while their incoming ties from former in-group members (now out-group members) are at high risk of being lost.

\section{Discussion}

As we have shown, inhomogeneous two-star terms are effective tools for representing stigmatic or other active group boundary maintenance processes within network models, capturing the tendency of group members to ostracize one of their own when they associate with a stigmatized out-group.  As they amplify the impact of differential mixing, it is reasonable to suspect that these two types of group maintenance effects will often co-occur in practice.  Indeed, we see exactly this in the context of the Parker and Asher study, which shows a combination of simple mixing inhibition and active maintenance.  It is plausible that this effect has been missed in prior studies of homophily, and testing for possible inhomogeneous two-star effects is something that should be considered where it is plausible that stigmatic or related processes are at work.  In such contexts, we would note that a \emph{lack} of inhomogeneous two-star effects may be just as interesting as their presence: their absence may suggest that homophily is driven either by ``neutral'' mechanisms like differential contact, or by strictly positive selection for alters with similar attributes.  

As the effects studied here are extensions of the conventional 2-star statistic, it is reasonable to consider whether they pose risks of degeneracy.  Degeneracy in ERGMs refers to a phenomenon that arises from unrealistic model specifications, in which probability becomes heavily concentrated on a small number of (usually implausible) structures as graph size increases. Degenerate models are informative, in that they indicate that the theory embodied by the proposed model leads to emergent network structures that are not realistic; given that one is attempting to model an observed system, however, avoiding degenerate models is a natural goal.\footnote{That said, if a seemingly plausible model for a given network turns out to be degenerate, one should view this as theoretically meaningful, just as in other cases in which a proposed model fails to explain one's data.}  Previous explorations of standard k-star terms have shown that (when not balanced as in alternating k-star sequences \cite{hunter:sn:2007}) they typically lead to degenerate distributions when paired with positive parameter values \cite{handcock:ch:2003}.  When $\theta>0$, the terms used here can pose a similar risk of degenerate behavior.  Intuitively, this can be understood as follows: when $\theta>0$, the more ties that an individual has to out-group members, the more attractive he or she is to in-group members; and, likewise, the more strongly an individual is tied to his or her in-group, the greater his or her tendency to have out-group ties.  This putative mechanism (if not checked by other effects) can lead to a positive feedback loop where boundary spanning encourages more in-group attention, which in turn feeds more boundary spanning, until a density explosion \cite{butts:sm:2011b} occurs.  This is implausible for a typical social network.

%\ctb{Note why degeneracy is unlikely to be a problem in the standard $\theta<0$ case.}
When $\theta<0$, however, our mechanism is very different: out-group ties tend to inhibit in-group ties (and vice-versa), leading to self-limiting behavior.  In this regime, inhomogeneous two-star terms cannot fuel a density explosion (because they do not enhance tie formation), and even their potential inhibitory effects are limited by bounds on mean degree (since penalties to tie formation scale with incoming and/or outbound ties).  Thus, in the regime of interest here, model degeneracy is not a major concern.

As with the general case of the 2-star statistic, which is a special case of the more general $k$-star family, it is possible to construct higher-order versions of the statistics described here.  A collection of such terms could subsequently be employed in curved statistics, analogous to the geometrically weighted$ $k-stars often employed to model degree distributions \cite{hunter:sn:2007}.  Such statistics could facilitate the modeling of diminishing marginal effects of boundary spanning on tie formation, rather than the constant incremental effect of the inhomogeneous two-star.  However, because inhomeneous stars of order greater than 2 can also vary in their composition, determining how stars of varying inhomogeneity should be weighted is non-trivial.  This would seem to be a potentially fruitful direction for subsequent work.

The network perturbation framework we introduced is not limited to calculating expected degree change and can be generalized to any network statistic of interest. Given the cross-sectional nature of the data and the assumptions involved, interpreting the differences in expected degree as dynamic predictions must be viewed as heuristic (at best); however, such ``what if'' experiments provide important insights into the behavior of structural mechanisms that would be difficult to obtain otherwise.  In particular, here they allow us to see that, for the Parker and Asher classroom, tie-forming dyadic and triadic closure mechanisms are outweighed by the effects of ostracism, and, ceteris paribus, perturbations leading ego to span group boundaries (whether by extending nominations or by changing their own status) would proximately be predicted to be associated with loss of friendship nominations.  Comparative analysis of such effects, combined with ethnographic or perceptual data, could be useful in assessing the extent to which individuals' perceived pressure to conform to group norms or refrain from cross-group interaction in fact correlate with the plausible risks of deviation.

%Every pair of ties that you have, to build out to gwdegree, diminishing marginal effects, alternating signs on them, each additional tie exerts more social pressure.
%Furthermore, while we have only considered the inhomogeneous 2-star in this paper, there is potential to extend this to an inhomogeneous k-star formulation. This would lead one to quite naturally formulate a geometrically weighted ingroup-degree term where the bias significant enough.
\section{Conclusion}
The ERGM framework provides a powerful framework to study the effects of multiple social processes in shaping a network structure. The inhomogeneous two-star term provides a simple and elegant means of incorporating active maintenance of group boundaries implied e.g. by some sociological theories of stigma easily into model-based network analysis.  Our simulation analyses demonstrate that this family of mechanisms operates differently from (and interacts with) simple differences in mixing rates, and our illustrative empirical analysis shows that the mechanisms are operative in at least some settings.  It is hoped that terms like that introduced here will serve to broaden the reach of parametric network modeling, and clarify the connections between structural biases and the social mechanisms that generate them.
%\ctb{Update.} 

%We have found that the coefficient associated with our inhomogeneous two-star is negative. When this statistic is negative, we would expect not only that if a node is heavily tied to a group that matches their ascriptive characteristic that there is an overall pressure, but we would also expect that if an ego is tied to an outgroup, that the people in ego's group won't want to tie to ego. When this statistic is positive, we expect that the more an ego is tied to an ingroup, the more that ego like to create ties to an outgroup. Apart from being qualitatively strange, we strongly suspect that this would result in ERGM degeneracy. Thus, we theorize that in most realistic social settings where this term may be used, that the coefficient associated with this term should either trend towards 0 or be negative.

It should be noted, that the strength of this approach for evaluating stigma in a population does not evaluate the respondent's level of perceived stigmatization. Rather, it evaluates the level of \textit{structural} stigma at the network level. Considering this can be evaluated at both the ego-level and the complete network level, this provides a tremendous amount of utility, both in understanding an ego's propensity to stigmatize, as well as a global metric of stigma. The achievement of evaluating stigma without attitudinal measures cannot be understated. First, from a data collection perspective, the only relevant questions that are needed are the respondents network ties and the relevant exogenous covariates of the ego and the alter, potentially reducing respondent fatigue. Second, it circumvents some issues of respondent (in)accuracy in data collection. If an ego is primed to present themselves as a person who does not stigmatize others, the alters they name may be deliberately cultivated in order to prove their open-mindedness. Furthermore, if the sociological phenomenon of interest is some behavioral change dependent on one's social ties (i.e. an influence process or contagion process), then it is irrelevant whether or not the ego perceives themselves as stigmatizing others, the relevant mechanism is dependent on the propensity for them to form ties across group and how that propensity is affected by the ties to their own group. Thus, this measure provides a more direct way of understanding how a stigmatic process affecting network structure can subsequently impede diffusion across group.

We conclude that this statistic can be powerful in explaining how cohesive subgroups based on some ascriptive characteristic (i.e. gender, race, etc.) arise from tendencies towards discrimination as opposed to pressures towards homophily. We present this as an important advance for the conceptualization of xenophobia on the network level.

\bibliography{las_compare1}

\begin{thebibliography}{}

\bibitem[Blau, 1994]{blau1994structural}
Blau, P.~M. (1994).
\newblock {\em Structural contexts of opportunities}.
\newblock University of Chicago Press.

\bibitem[Butts, 2007]{butts:sm:2007b}
Butts, C.~T. (2007).
\newblock Models for generalized location systems.
\newblock {\em Sociological Methodology}, 37(1):283--348.

\bibitem[Butts, 2011]{butts:sm:2011b}
Butts, C.~T. (2011).
\newblock Bernoulli graph bounds for general random graphs.
\newblock {\em Sociological Methodology}, 41:299--345.

\bibitem[Goffman, 1963]{goffman:bk:1963}
Goffman, E. (1963).
\newblock {\em Stigma: Notes on the Management of Spoiled Identity}.
\newblock Simon and Schuster, New York.

\bibitem[Goodreau et~al., 2017]{goodreau2017sources}
Goodreau, S.~M., Rosenberg, E.~S., Jenness, S.~M., Luisi, N., Stansfield,
  S.~E., Millett, G.~A., and Sullivan, P.~S. (2017).
\newblock Sources of racial disparities in hiv prevalence in men who have sex
  with men in {A}tlanta, {GA}, {USA}: a modelling study.
\newblock {\em The Lancet HIV}, 4(7):e311--e320.

\bibitem[Handcock, 2003]{handcock:ch:2003}
Handcock, M.~S. (2003).
\newblock Statistical models for social networks: Inference and degeneracy.
\newblock In Breiger, R., Carley, K.~M., and Pattison, P., editors, {\em
  Dynamic Social Network Modeling and Analysis}, pages 229--240. National
  Academies Press, Washington, DC.

\bibitem[Handcock et~al., 2008]{handcock.et.al:jss:2008}
Handcock, M.~S., Hunter, D.~R., Butts, C.~T., Goodreau, S.~M., and Morris, M.
  (2008).
\newblock {statnet}: Software tools for the representation, visualization,
  analysis and simulation of network data.
\newblock {\em Journal of Statistical Software}, 24(1):1--11.

\bibitem[Heider, 1946]{heider:jp:1946}
Heider, F. (1946).
\newblock Attitudes and cognitive organization.
\newblock {\em Journal of Psychology}, 21:107--112.

\bibitem[Hunter, 2007]{hunter:sn:2007}
Hunter, D.~R. (2007).
\newblock Curved exponential family models for social networks.
\newblock {\em Social Networks}, 29:216--230.

\bibitem[Hunter et~al., 2013]{hunter.et.al:jss:2013}
Hunter, D.~R., Goodreau, S.~M., and Handcock, M.~S. (2013).
\newblock ergm.userterms: A template package for extending statnet.
\newblock {\em Journal of Statistical Software}, 52(2):1--25.

\bibitem[Hunter et~al., 2008]{hunter.et.al:jss:2008}
Hunter, D.~R., Handcock, M.~S., Butts, C.~T., Goodreau, S.~M., and Morris, M.
  (2008).
\newblock ergm: A package to fit, simulate and diagnose exponential-family
  models for networks.
\newblock {\em Journal of Statistical Software}, 24(3).

\bibitem[Krackhardt and Stern, 1988]{krackhardt.stern:spq:1988}
Krackhardt, D. and Stern, R.~N. (1988).
\newblock Informal networks and organizational crises: An experimental
  simulation.
\newblock {\em Social Psychology Quarterly}, 51:123--140.

\bibitem[McPherson et~al., 2001]{mcpherson2001birds}
McPherson, M., Smith-Lovin, L., and Cook, J.~M. (2001).
\newblock Birds of a feather: Homophily in social networks.
\newblock {\em Annual review of sociology}, 27(1):415--444.

\bibitem[Parker and Asher, 1993]{parker1993friendship}
Parker, J.~G. and Asher, S.~R. (1993).
\newblock Friendship and friendship quality in middle childhood: Links with
  peer group acceptance and feelings of loneliness and social dissatisfaction.
\newblock {\em Developmental psychology}, 29(4):611.

\bibitem[Sakoda, 1971]{sakoda:jms:1971}
Sakoda, J.~M. (1971).
\newblock The checkerboard model of social interaction.
\newblock {\em Journal of Mathematical Sociology}, 1(1):119--132.

\bibitem[Schaefer et~al., 2017]{schaefer2017friends}
Schaefer, D.~R., Bouchard, M., Young, J.~T., and Kreager, D.~A. (2017).
\newblock Friends in locked places: an investigation of prison inmate network
  structure.
\newblock {\em Social networks}, 51:88--103.

\bibitem[Schelling, 1971]{schelling1971dynamic}
Schelling, T.~C. (1971).
\newblock Dynamic models of segregation.
\newblock {\em Journal of mathematical sociology}, 1(2):143--186.

\end{thebibliography}
\end{document}